# Fano-type effect in hydrogen-terminated pure nanodiamond

Oleg S. Kudryavtsev[1], Rustem H. Bagramov[2], Arkady M. Satanin[3,4], Andrei Shiryaev[5], Oleg I. Lebedev[6], Alexey M. Romshin[1], Dmitrii G. Pasternak[1], Alexander V. Nikolaev[7,8], Vladimir P. Filonenko[2], Igor I. Vlasov[1*]

[1]Prokhorov General Physics Institute of the Russian Academy of Sciences, 119991 Moscow .Russia
[2]Vereshchagin Institute of High Pressure Physics, Russian Academy of Sciences, 108840 Moscow, Troitsk, Russia
[3]Dukhov All-Russia Research Institute of Automatics, 101000 Moscow, Russia
[4]National Research University Higher School of Economics, 101000 Moscow, Russia
[5]Frumkin Institute of Physical Chemistry and Electrochemistry RAS, 119071 Moscow, Russia
[6]Laboratoire CRISMAT, UMR 6508 CNRS-ENSICAEN, 14050, Caen, France
[7]Skobeltsyn Institute of Nuclear Physics Lomonosov Moscow State University, 119991 Moscow, Russia
[8]School of Electronics, Photonics and Molecular Physics, Moscow Institute of Physics and Technology, 141700 Dolgoprudny, Moscow region, Russia

*- E-mail: vlasov@nsc.gpi.ru

**Two novel properties, unique for semiconductors: a negative electron affinity [1-2], and a high p-type surface electrical conductivity [3-4], were discovered in diamond at the end of the last century. Both properties appear when the diamond surface is hydrogenated. A natural question arises: is the influence of the surface hydrogen on diamond limited only to the electrical properties? Here, we report the first observation of a transparency peak at 1328 cm$^{-1}$ in IR absorption of hydrogen-terminated pure (undoped) nanodiamonds. This new optical property is ascribed to Fano-type destructive interference between zone-center phonons and free carriers (holes) appearing in the near-surface layer of hydrogenated nanodiamond. Our work opens the way to exploring the physics of electron-phonon coupling in undoped diamonds and promises the application of the H-terminated nanodiamonds as a new optical material with an induced transparency in IR optical range.**

The surface electrical conductivity in diamonds is explained using the model of electrochemical "transfer doping", in which an electron transfer occurs from the hydrogen terminated diamond surface to the adsorbed water layer, resulting in hole accumulation and an upward band bending of the valence band maximum near the diamond's surface [4]. Transfer doping of diamond generates hole carriers with typical concentration of $10^{12}$–$10^{13}$ cm$^{-1}$ [5,6]. Taking into account that a typical width of near surface layers does not exceed 1 nm [7], one obtains a relatively high volume concentration of free charge carriers $10^{19}$-$10^{20}$ cm$^{-3}$. It is known that at such high concentrations of free carriers in doped semiconductors one can observe Fano-type interference [8,9] between the discrete states of optical phonons and continuum of electron (hole) levels induced by substitutional impurities. This interference manifests itself in an asymmetric profile of phonon lines in Raman spectra of heavily doped semiconductors, including boron-doped diamonds [10-14]. Fano effect has been observed earlier in IR absorption for silicon and germanium containing a large concentration of acceptor impurities [15,16,17]. It is natural to suggest analogous Fano-type interference in the near surface region of highly hydrogenated diamond, containing a large concentration of free carriers. In absence of dopants its lattice remains perfect ensuring high Q-factor of the phonon oscillator and, subsequently, enhancing the effect of Fano interference. The choice of nano-sized diamond possessing the increased surface-to-volume ratio creates favorable conditions for the experimental detection of the Fano interference. Nevertheless, so far there has been no reports on the electron-phonon interaction neither in Raman scattering nor in IR absorption of hydrogenated NDs. The main reasons for lack of such works are assumed to be widespread use of highly defective detonation nanodiamonds, containing a large amount of nitrogen impurity and structural defects, as well as difficulties in efficient surface hydrogenation by thermochemical treatment [18].

In the present work, to detect Fano-type interference we employ a new generation of nanodiamonds produced from hydrocarbons under high pressure and high temperature (HPHT) (see Methods). As has been shown earlier [19,20] and confirmed in the present work, the surface of ND particles synthesized from hydrocarbon compounds is effectively terminated by hydrogen. This is evidenced by the observation of $CH_x$ phonon modes in Raman, while they are never observed in Raman of NDs hydrogenated using thermochemistry.

Diamond crystallite sizes were determined by Transmission Electron Microscopy (TEM) (Fig.1a) and X-ray diffraction analysis (Fig.1S in SI, Section S1). TEM studies show that characteristic sizes of diamond nanocrystals are varying in the range of 20-40 nm (Fig.1a), although crystals ~1 micron were occasionally encountered. Many diamonds tend to twin along the <111> plane (Fig.1b,c). High resolution TEM images show a perfect lattice, {111} morphology (Fig 1c), sharp and clean surface of diamond crystallites (Fig1c insert).

The crystal sizes of near 30 nm are supposed to be optimal for finding out the Fano-type interference in hydrogenated diamond. Smaller particles appear less promising, since they can have weaker negative electron affinity and lower near-surface hole concentration than chosen ones, as shown by A. Bolker et al. [6]. Note high secondary electron yield of the NDs manifested as unusually bright SEM image of the as-grown NDs (Fig. S2 in SI, Section S2) caused by negative electron affinity.

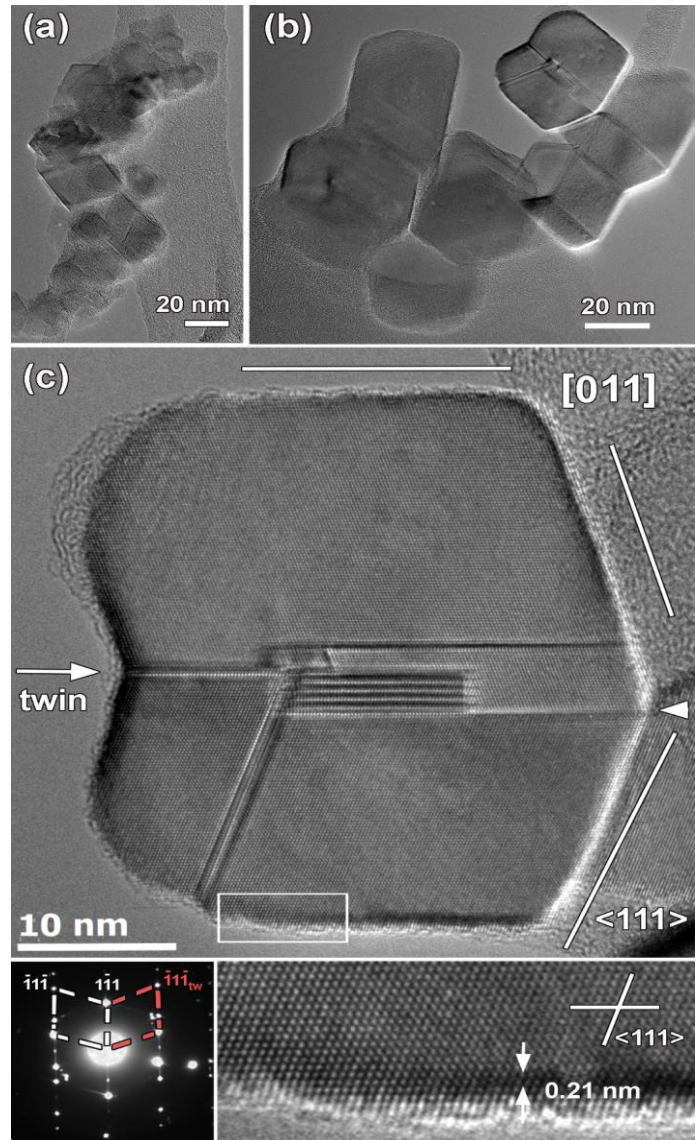

Fig. 1. Transmission electron images of the nanodiamonds. (a) - BF low magnification overview TEM image of ND sample (b) - low magnification TEM image of highly twinned NDs (c) – [011] HRTEM image of one of the twinned crystallites shown in (b). Corresponding twin ED pattern is shown in the left insert. Matrix and twin spots are marked with white and red boxes, respectively. The magnified image of the near-surface region marked in (c) is shown in the right insert.

Raman and IR absorption spectroscopies were used to find out the Fano-type interference in the ND sample.

In the Raman spectrum a slight asymmetry in the diamond line profile centered at 1332 cm$^{-1}$ is observed (Fig.2a). This asymmetry is definitely associated with the Fano interference between optical phonons and hole states in the near-surface diamond layer. However, for 30-nm diamond particles the contribution of this layer is relatively small, and the interference is masked by the intense symmetric peak of optical phonons arising in diamond bulk. Thus, a detailed analysis of expected interference is not possible by Raman spectroscopy.

On annealing the sample in air at 400 $^0$C (see Methods) a major fraction of hydrogen is removed from the ND surfaces (see SI), and the resulting diamond line is well approximated by the symmetric Lorentz profile (Fig. 2c). This experiment confirms that the profile asymmetry is related to the hydrogen-terminated ND surface.

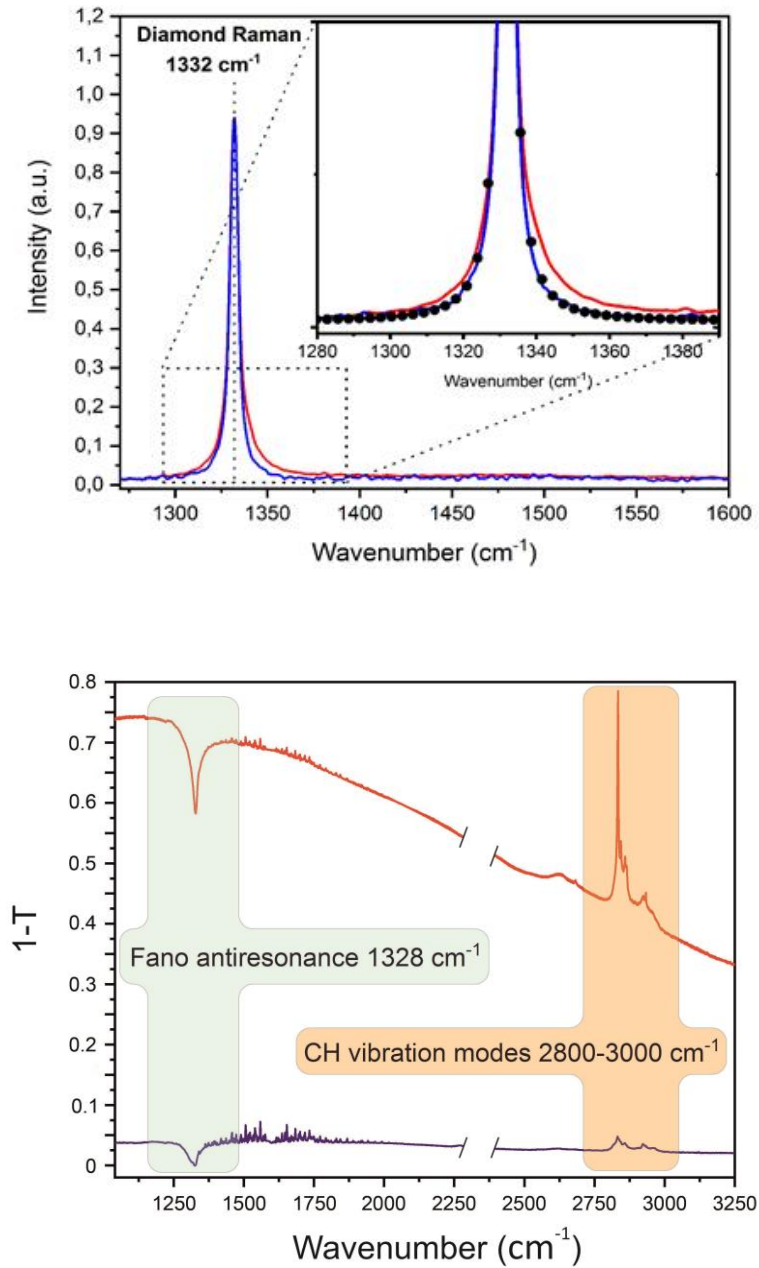

Fig. 2 Raman and infra-red spectra of the sample. (a) - Room temperature Raman spectra of the initial (red line) and annealed at 400 $^0$C (blue line) samples. The spectra are characterized by a narrow diamond line at 1332 см$^{-1}$. The inset demonstrates asymmetry of the red line. After annealing the diamond line is well fitted by the Lorentz profile indicated by black dots. (b) - IR absorption spectra (in units 1-T, where T is transmittance) measured at room temperature for the initial (red line) and annealed at 400 $^0$C (blue line) samples. The spectra are corrected for the scattering losses at 6000 cm$^{-1}$, which are assumed to be constant in the range 1000-6000 cm$^{-1}$. A narrow dip (transparency peak) at 1328 cm$^{-1}$ is related to the Fano-type destructive interference between phonons and holes. Omitted region around 2370 cm$^{-1}$ contains uncompensated atmospheric $CO_2$ absorption. Structured band within 2800-3000 cm$^{-1}$ is due to $CH_x$ vibration modes. Background absorption, steadily increasing towards low frequencies, is attributed to free carriers in diamond. Note a considerable decrease of the 1328 cm$^{-1}$ dip and marked decrease of the background after removal of the main fraction of hydrogen from the ND surface.

IR absorption spectroscopy turned out much more favorable than Raman for observing the Fano-type interference in hydrogenated nanodiamonds. According to the selection rules [21] pure diamond does not absorb IR radiation in the frequency range of the zone-center optical phonons. As a result, a surface of hydrogenated diamond becomes the main source of IR absorption in this range.

The IR absorption spectrum of our sample is shown in Fig. 2b. Its main characteristics are:
- the narrow dip in absorption (transparency peak) at frequency 1328 cm$^{-1}$,
- the intense band of CH$_x$ vibration modes in the range 2800-3000 cm$^{-1}$,
- the background absorption steadily increasing towards low frequencies (starting from 6000 cm$^{-1}$).

Absence of characteristic C-F bond absorption within 1100-1300 cm$^{-1}$ [22] implies absence of contamination of diamond surface by fluorine from the precursor.

On annealing the sample in air at 400 $^0$C (see Methods) an substantial drop of the CH$_x$ absorption is accompanied by a decrease in the wide-band absorption and the transparency peak. These changes demonstrate the relation of the transparency peak and the wide-band absorption to hydrogen, which terminates the ND surface.

Now we turn to interpretation of the main features detected in the absorption spectrum of the pristine sample. A wide-band absorption we attribute to the continuum of hole states, and the transparency peak at 1328 cm$^{-1}$ – to the destructive interference between the holes and zone-center optical phonons in diamond. The closest analogue of the observed phenomenon could be coupled-resonator-induced transparency (CRIT) [23-26]. In CRIT the destructive interference is the result of an internal coupling between individual resonators. In our case we deal with phonon-hole coupling. The important condition for the CRIT effect manifestation is a difference in losses of coupled resonators. The CRIT resonator non-interacting with the external field should have considerably smaller losses in comparison with the interacting one. Similarly, the Q-factor of the phonon oscillator was demonstrated to be much higher than the hole one (see SI, Section S5).

Below we describe the nature of the phonon-hole interference in terms of "phonon" and "hole" oscillators. The hole oscillator is described in terms of the plasmonic model (SI) with the amplitude $A_h$, the eigenfrequency $\omega_h$ and the damping coefficient $\gamma_h$. This oscillator is excited by an incoming electromagnetic field with the amplitude $A_{ext}$. The optical phonon oscillator is described within the Einstein model (SI) with the amplitude $A_p$, the eigenfrequency $\omega_p$, and the damping coefficient $\gamma_p$. As we mentioned above, $\gamma_p \ll \gamma_h$. In the range of resonance frequencies, $\omega \sim \omega_h : \omega_p$, the approximate equations for the amplitudes of the coupled oscillators in the external field (see details in SI, Section S4) are given by:

$$(\delta_h + i\gamma_h/2)A_h + \bar{k}A_p = A_{in}, \qquad (1)$$
$$\bar{k}A_h + (\delta_p + i\gamma_p/2)A_p = 0,$$

where $\bar{k} = k/2\omega_p$ and $k$ is the coupling coefficient of the oscillators, while $\delta_h = \omega - \omega_h$ and $\delta_p = \omega - \omega_p$ are the frequency detuning parameters.

Expressing the dielectric susceptibility in terms of the amplitude ratio $\frac{A_h}{A_{in}}$ and solving Eq. (1) in the limit $\gamma_p \to 0$ we obtain the following dependence of the imaginary part of the dielectric susceptibility $Im\chi(\omega)$, describing the absorption of two interacting oscillators, on the external field frequency:

$$Im\chi(\omega) = \frac{e^2 n}{2\omega_p m_h \varepsilon_0} \frac{\gamma_h/2}{(\gamma_h/2)^2 + (\delta_h - \bar{k}^2/\delta_p)^2}, \qquad (2)$$

where $n$ is a hole concentration in near-surface diamond layer, $m_h$ - hole mass.

The $Im\chi(\omega)$ profiles are analyzed for 3 main regimes:

i) The oscillators are uncoupled, $k = 0$, then $Im\chi(\omega)$ is described by a wide Lorenz profile, shown in Fig. 3a. The phonon oscillator is completely decoupled from the hole oscillator and forms the so-called "dark state".

ii) The oscillators are weakly coupled, i.e. $k$ is small, but not equal to zero. In that case, when $\delta_p \rightarrow 0$, $Im\chi(\omega) \rightarrow 0$ and a sharp dip appears in the wide absorption profile at $\omega_p$ (Fig. 3b). This regime is most consistent with our experimental observations.

iii) In the strong-coupling limit ($k \gg 0$) the absorption profile splits into two bands strongly separated in frequency. It means that in the absence of damping, the eigenmodes of the oscillators strongly depend on the coupling constant $k$. We believe that such strongly coupled polaritonic excitations cannot be realized in non-polar covalent crystals.

The model proposed here for explanation of the dip in absorption is simplified, but it reflects the main features of Fano-type destructive interference. The fitting of the experimental dip in the absorption spectrum (see SI, Section S4) determines its position at 1328 cm$^{-1}$ and the width of 30 cm$^{-1}$. The depth of the dip is 1/5 of the ND absorption near the dip, as follows from Fig. 2b. Numerical analysis of the expressions for dielectric susceptibility (2) and more general (7S), obtained at $\gamma_p \neq 0$, shows that the width of the dip depends mainly on the coupling coefficient $k$, whereas the depth of the dip - on the value of phonon damping coefficient $\gamma_p$. On strengthening the coupling between oscillators the dip broadens and weakens with increasing, respectively, $k$ and $\gamma_p$. Our investigation of the dip characteristics as a function of a nanodiamond size is in progress.

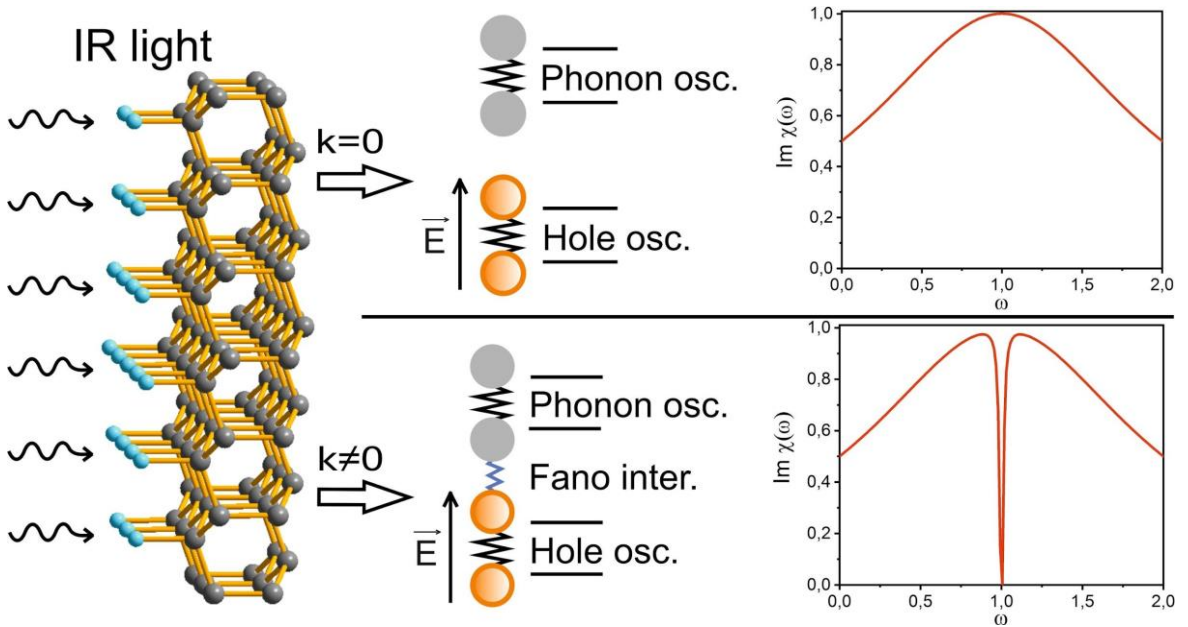

Fig. 3. Qualitative picture of the phonon-hole interference in H-terminated diamond. Excitation of hole oscillators by incoming electromagnetic field E is shown schematically.
(a) When the phonon and hole oscillators are decoupled, the Lorentz absorption profile is observed due to hole excitations, while the phonon oscillator remains "dark".
(b) Taking into account the coupling between the oscillators, the phonon oscillator compensates for dissipation losses of the hole oscillator in a narrow frequency range near the optical phonon frequency, causing a narrow dip in the Lorentz absorption profile.

In conclusion, we have found a narrow dip at the frequency 1328 cm$^{-1}$ in IR absorption spectra of 30-nm H-terminated diamond particles synthesized at high pressure. We relate this effect, unusual for pure diamond, to the destructive interference of the Fano-type between zone-center phonons and charge carriers appearing in the near-surface layer of the hydrogenated nanodiamond. The destructive interference is explained by a simple model of two coupled oscillators, one of which interacts with the external electromagnetic wave. Detected phonon-hole coupling [27] in hydrogen-terminated nanodiamond allow us to speculate that, when reaching (if possible) high local concentrations of holes, such NDs could be superconducting similar to diamond doped with boron at the level >$10^{21}$ cm$^{-3}$ [28]. A sharp dip in absorption near the optical phonon frequency gives rise to an anomalous light dispersion in this region. Materials with this property are able to reduce the group velocity of the transmitted light [29]. Thus, the hydrogenated nanodiamond represents a new optically active media which can be used for controlling the velocity of IR light pulses. "Slowed" light is currently employed for the implementation of the optical buffer [30], for quantum networks [31], and for designing the quantum memory [32]. On the whole, finding the IR transparency peak induced by phonon-hole coupling opens a new page in investigation and application of diamond in IR optical range.

**Methods**
**1. Nanodiamonds synthesis and sample preparation**

Nanodiamonds were synthesized from a mixture of adamantane ($C_{10}H_{16}$, Aldrich, 99% purity) and octafluoronaphthalene ($C_{10}F_8$, Alpha Aesar, 96% purity), which were manually mixed in an agate mortar (the octafluoronaphthalene/adamantane weight ratio was 1/4). The mixture was placed in a titanium capsule and then in a graphite heater inside a lithographic stone container. For treatment at 7.5 GPa and 1400 $^0$C, apparatus of toroidal type was used [33]. The synthesized NDs is a white powder with bluish hint. Due to their hydrophobicity, ND particles are poorly dispersed in water and upon drying of suspension on a substrate they form agglomerates of various sizes.

**2. Sample annealing**
Samples were annealed in the air in the LinKam TS1500 chamber for 30 minutes at 400 $^0$C. The heating rate was 120 $^0$C/minute. Raman and IR absorption spectra were recorded after cooling to room temperature.

**3. Raman spectroscopy**
Raman spectra of the nanodiamonds were recorded at room temperature with a LABRAM HR800 spectrometer equipped with a 473-nm diode laser. The exciting laser radiation of the power 10 mW was focused on samples by means of the 100x Olympus lens and NA=0.95. The scattered radiation was registered in the backscattering geometry.

**4. IR absorption spectroscopy**
IR absorption spectra of the nanodiamonds were recorded at room temperature using
IR microscope Thermo Fisher Scientific Nicolet iN10. The spectra were recorded in the mid-infrared range 1000–6000 см$^{-1}$ with a resolution of 2 см$^{-1}$ . In the absorption mode the samples were dispersed on a KBr pellet and measured with square apertures ranging from 30 to 150 µm depending on size of the ND agglomerates. In addition, reflectance spectra from samples dispersed on Al or Au mirrors were studied. As expected, the 1328 cm$^{-1}$ feature changes sign in the reflectance spectra.

## 5. TEM analysis

Transmission electron microscopy (TEM) was performed using JEM ARM200F cold FEG double aberration corrected microscope, operated at 200 kV and equipped with a large angle CENTURIO EDX detector, Orius Gatan CCD camera and Quantum GIF. The TEM samples were prepared in a conventional way – depositing a solution of the material on a carbon holey Cu grid.

# Supplementary information

**S1. X-ray diffraction (XRD) analysis**

X-ray diffraction pattern was acquired at Empyrean diffractometer (Panalytical BV) using Ni-filtered Cu Kα radiation in Bragg-Brentano (reflection) geometry. The sample were measured on zero-background Si holder.

XRD phase analysis of the ND sample reveals the effective transformation of the initial hydrocarbons into diamond without any noticeable formation of the graphite phase of carbon (Fig. S1). Detailed examination of diamond reflections suggest the presence of two fractions of diamond grains – small and large (~1 micron) ones. Williamson-Hall plot indicates that the small grains are approx. 30 nm in size. Despite the low amount of large grains (see main text), they make a significant contribution to X-ray diffraction due to their big volume.

The presence of a small amount of crystalline $CaF_2$ in the sample (Fig. S1) we associate with the reaction of fluorine (released during the thermobaric decomposition of octafluoronaphthalene) with calcium, which is part of the chamber material.

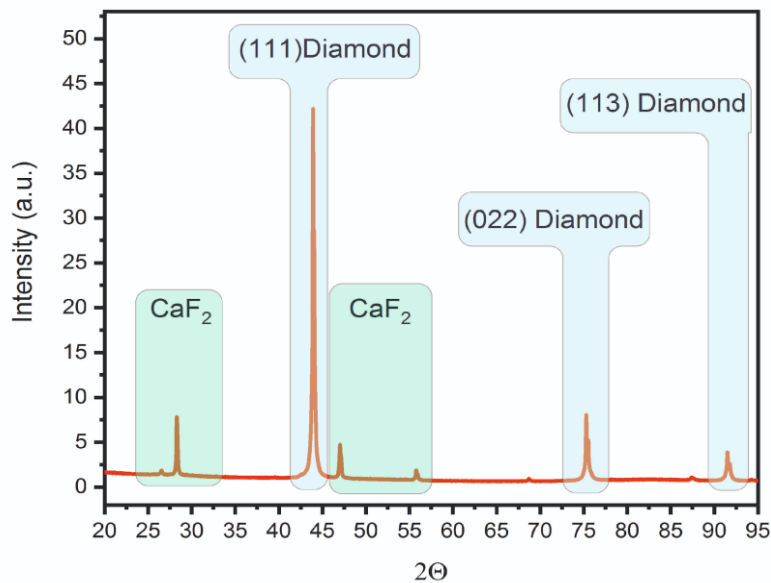

**Figure S1.** X-ray diffraction pattern of H-terminated nanodiamonds

**S2. Scanning electron microscopy (SEM)**

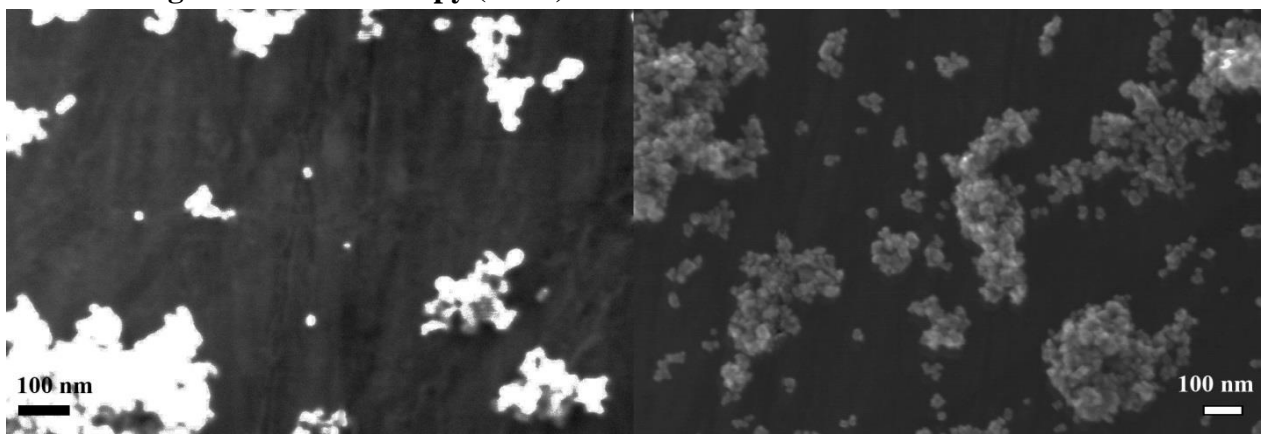

**Figure S2.** SEM images of H-terminated nanodiamonds before and after annealing. The right image is typical for oxidized NDs, whereas is unusually bright left image is explained by strong back scattering of secondary electrons from the negative-electron-affinity surface of H-terminated NDs.

## S3. Raman spectra of C-H vibration modes.

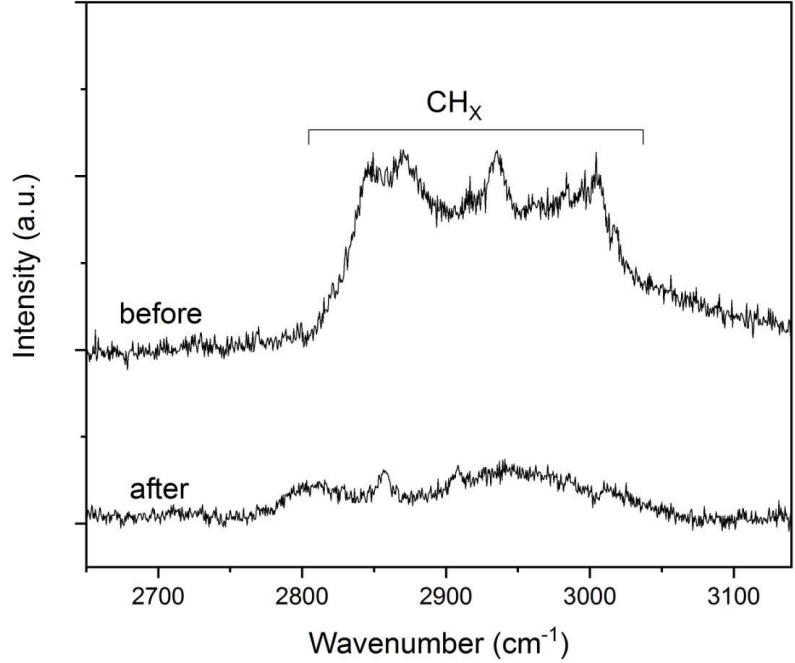

**Figure S3**. Raman spectra of C-H$_x$ vibration modes before and after annealing. A significant weakening of CH$_x$ bands indicates the effective removal of hydrogen from the nanodiamonds surface.

## S4. Phonon-hole interference

To substantiate the manifestation of the Fano-type destructive interference in our experiments, we consider a simple scenario demonstrating the physical nature of the interference between the optical phonons and hole carriers.

We start with expressing the absorbance shown in Fig. 2(a), through the imaginary part of the dielectric susceptibility. From Maxwell's equations it follows that the intensity $I_R(\omega)$ absorbed by a near-surface layer of the thickness $L$ is given by

$$I_R(\omega) = Im\chi(\omega) L k_{ir} I_0(\omega) \tag{s1}$$

where $I_0(\omega)$ is the intensity of the incident IR radiation, $Im\chi(\omega)$ is the imaginary part of the dielectric susceptibility, which will be found below within a simple model of coupled oscillators. In addition, in deriving of Eq. (s1) we assume that $Lk_{ir} \ll 1$ where $k_{ir}$ is the wave vector of the IR radiation. The condition $Lk_{ir} \ll 1$ implies that the wave field is weakly changed across the width of the layer, i.e. the absorption is weak.

The dielectric susceptibility is found from the equation connecting Fourier components of the polarization vector $P(\omega)$ with components of the external field $E(\omega)$: $P(\omega) = \varepsilon_0 \chi(\omega) E(\omega)$, where $\varepsilon_0$ is the dielectric constant. (Here the optical media is considered isotropic and therefore the polarization vector has only one component parallel to the external field. In addition, only the transverse components of fields are allowed.)

In the following we assume that in the frequency range of interest the polarization is caused by the action of two oscillators, one of which is the hole oscillator driven by the external field and the other is the phonon oscillator, which gets excited only through the coupling with the hole one.

The hole oscillator is described in terms of the plasmonic model with the eigenfrequency $\omega_h$ and the damping coefficient $\gamma_h$. At hole oscillations, from the condition of local electroneutrality, one finds the restoring local electric field $E = -\varepsilon_0 enx$. Adding to it a damping term and the driving force from the incident electromagnetic field, we obtain the following equation of motion:

$$\ddot{x} + \gamma_h \dot{x} + \omega_h^2 x = -\frac{e}{m_h} E_{in}(t) \tag{s2}$$

where $\omega_h = (e^2 n/m_h \varepsilon_0)^{1/2}$, and $m_h$ is the hole mass.

The optical phonon oscillator described within the Einstein model [1] is approximated by its eigenfrequency $\omega_p$ and the damping coefficient $\gamma_p$. The phonon-hole coupling makes possible the exchange of energy between the hole and phonon subsystems. The equations of motion for this case become

$$\ddot{x} + \gamma_h \dot{x} + \omega_h^2 x + ky = -\frac{e}{m_h} E_{in}(t), \tag{s3}$$
$$\ddot{y} + \gamma_p \dot{y} + \omega_p^2 y + kx = 0,$$

where $k$ is the coupling coefficient for two oscillators.

Substituting $E_{in}(t) = Ee^{-i\omega t} + c.c.$ (where $c.c.$ means complex conjugate) $E_{ext}(t) = Ee^{-i\omega t} + c.c.$, $x(t) = A_h e^{-i\omega t} + c.c.$ and $y(t) = A_p e^{-i\omega t} + c.c.$ in Eq. (s3) we obtain

$$(-\omega^2 + \omega_h^2 - i\omega\gamma_h)A_h + kA_p = -\frac{e}{m}E, \tag{s4}$$
$$kA_p + (-\omega^2 + \omega_p^2 - i\omega\gamma_p)A_p = 0.$$

Equalizers (s4) describes the coupled oscillations of a system over a wide frequency range. For greater clarity, we use the method of slowly varying amplitudes, which well describes the behavior of the system in the region near the dip and the resonances. In the resonant frequency region $\omega \sim \omega_h: \omega_p$, we additionally require the implementation of the inequalities $\omega = \omega_h + \delta_h$, $|\delta_h| \ll \omega_h$, $\delta_p = \frac{\omega_p^2 - \omega^2}{2\omega_p} + \delta_h$ and find the following equations for coupled amplitudes:

$$(\delta_h + i\gamma_h/2)A_h + \bar{k}A_p = A_{in}, \tag{s5}$$
$$\bar{k}A_h + (\delta_p + i\gamma_p/2)A_p = 0,$$

where $\bar{k} = k/2\omega_p$ and $A_h$, $A_p$ are the amplitudes of the hole and phonon oscillators, respectively, while $A_{in} = \frac{e}{2\omega_p m_h} E$ is the driving amplitude.

The Fourier component of polarization is the product of the dipole moment of the hole, $-eA_h$, and their concentration, $P(\omega) = -eA_h n$. Expressing $A_h$ from Eq. (s5) and comparing it with $P(\omega) = \varepsilon_0 \chi(\omega) E(\omega)$ at $\gamma_p = 0$, we arrive at

$$Im\chi(\omega) = \frac{e^2 n}{2\omega_p m_h \varepsilon_0} \frac{\gamma_h/2}{(\gamma_h/2)^2 + (\delta_h - k^2/\delta_p)^2} \tag{s6}$$

In the main text it is shown that $Im\chi(\omega)$ becomes zero when the frequency of the external field approaches the frequency of the phonon oscillator, $\delta_p = \omega - \omega_p \to 0$.

We now discuss the impact of the coupling constant $k$ and damping coefficient $\gamma_p$ on the shape and position of the dip in absorption.

From Eq. (s5) one generally finds

$$\chi(\omega) = -\frac{e^2 n}{2\omega_p m_h \varepsilon_0} \frac{(\delta_p + i\gamma_p/2)}{(\delta_h + i\gamma_h/2)(\delta_p + i\gamma_p/2) - \bar{k}^2} \tag{s7}$$

From the dispersion theory it follows that the anomalies in the transmission, absorption and reflection of the radiation are defined by zeros and poles of the dielectric susceptibility as a function of the frequency in the complex plane [1].

An inspection of Eq. (s7) shows that in the absence of dissipative processes for the phonon resonator, $\gamma_p = 0$, the dielectric susceptibility goes to zero at $\delta_p = \omega - \omega_0 = 0$, in accordance with the assumption made earlier and as it is follows from Eq. (s6).

If the oscillators are decoupled, $k = 0$, then the Eq. (s7) describes only one wide resonance due to the excitation of the hole oscillator. In this case, the expression (s 7) has only one pole at the frequency $\omega_h^{pole} = \omega_h - i\frac{\gamma_h}{2}$.

When $k \neq 0$ the poles of the dielectric susceptibility are defined by the equation $(\delta_h + i\gamma_h/2)(\delta_p + i\gamma_p/2) - \bar{k}^2 = 0$. Since the equation for the poles is of the second order, we get two poles for this case. It is easy to show that one of the poles approximately coincides with the previous expression, $\omega_h^{pole}$, which determines the absorption due to the hole oscillator.

A second pole appears as a solution of second order equation, for which we find approximately for small $k$ the expression

$$\omega_p^{pole} = \omega_p + \frac{\bar{k}^2 \delta_h}{\delta_h^2 + \left(\frac{\gamma_h}{2}\right)^2} - i\left(\frac{\gamma_p}{2} + \frac{\bar{k}^2 \frac{\gamma_h}{2}}{\delta_h^2 + \left(\frac{\gamma_h}{2}\right)^2}\right) \tag{s8}$$

The resulting expression shows that the interaction of the photon oscillator with the hole one gives the correction to the pole position and the resonance width proportional to $k^2$.

An important conclusion follows from the our analysis. Since the second pole is not visible in the experiment, this means that the contribution from the second pole, which lies in the complex plane, is completely shielded by zero in the absorption. This means that the coupling constant is really small, so there is only a dip Fano in the system under the consideration.

The above conclusions also mean that for the interpolation of experimental data, we can choose the fitting function in the form:

$$Im\chi(\omega) = C\frac{(\omega-\omega_p)^2}{(\omega-\omega_p)^2 + (\Gamma/2)^2} \tag{s9}$$

where C and $\Gamma$ are fitting parameters.
The results of the fitting are given in Fig. 4S. The width of the dip is approximately 30 cm$^{-1}$.

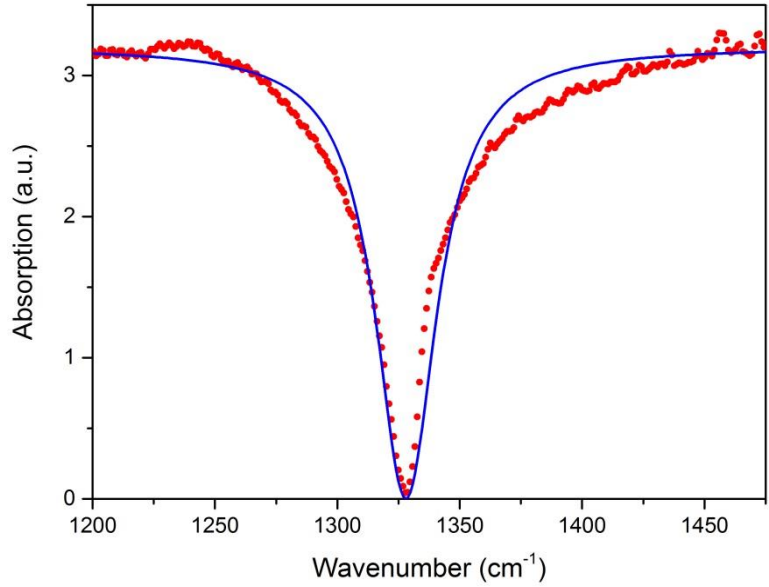

**Fig. S4.** The profile of the dip in absorption: points – experimental values, solid line – the result of the fit using Eq. S9. The background level interpolated by a linear function in the frequency range from 1200 cm$^{-1}$ to 1600 cm$^{-1}$ has been subtracted.

## S5. Q-factor estimation for the phonon and hole oscillators.
The hole oscillator:
To estimate the frequency of holes in the Fermi gas, we use an approximate expression from [2]:
where and is, respectively, the temperature and chemical potential of the electron gas (the proportionality coefficient in this expression is of the order of 1). For the estimation of $\omega_h$ and $\mu$ the characteristic concentration of holes n=1x10$^{19}$ cm$^{-3}$ in the near-surface layer of H-terminated diamond is taken, then $\omega_h = 0.21$ eV and $\mu = 0.05$ eV, and the hole oscillator quality factor ≈18.
The phonon oscillator:
A width of Raman line of 5 cm$^{-1}$ corresponds to a phonon dephasing time T = 1/$\Gamma$ of 2.2 ps and the phonon oscillator quality factor $Q_p = T\omega_p \approx 550$ [3].